\documentclass[%
 aip,
 amsmath,amssymb,
 reprint,%
]{revtex4-1}

\usepackage{graphicx}
\usepackage{dcolumn}
\usepackage{bm}

\usepackage[utf8]{inputenc}
\usepackage[T1]{fontenc}
\usepackage{mathptmx}
\usepackage{etoolbox}

\usepackage{color}

\newcommand{\nfp}{n_{\mathrm{fp}}}
\usepackage{hyperref}

\makeatletter
\def\@email#1#2{%
 \endgroup
 \patchcmd{\titleblock@produce}
  {\frontmatter@RRAPformat}
  {\frontmatter@RRAPformat{\produce@RRAP{*#1\href{mailto:#2}{#2}}}\frontmatter@RRAPformat}
  {}{}
}%
\makeatother
\begin{document}


\title[Stellarator optimization for good magnetic surfaces at the same time as quasisymmetry]{Stellarator optimization for good magnetic surfaces at the same time as quasisymmetry}
\author{Matt Landreman}
 \email{mattland@umd.edu}
\affiliation{ 
Institute for Research in Electronics and Applied Physics, University of Maryland, College Park MD 20742, USA
}%

\author{Bharat Medasani}
\author{Caoxiang Zhu}
\affiliation{%
Princeton Plasma Physics Laboratory, P.O. Box 451, Princeton NJ 08543, USA
}%

\date{\today}

\begin{abstract}
A method is demonstrated to optimize a stellarator's geometry to eliminate magnetic islands and achieve other desired physics properties at the same time.
For many physics quantities that have been used in stellarator optimization, including quasisymmetry, neoclassical transport, and magnetohydrodynamic stability, it is convenient to use 
a magnetic equilibrium representation that assures the existence of magnetic surfaces.
However, this representation hides the possible presence of magnetic islands, which are typically undesirable.
To include both surface-based objectives and island widths in a single optimization,
two fixed-boundary equilibrium calculations are run at each iteration of the optimization: one that enforces the existence of magnetic surfaces (VMEC [S. P. Hirshman and J. C. Whitson, Phys. Fluids \textbf{26}, 3553 (1983)]), and one that does not (SPEC [S. R. Hudson,  \textit{et al}, Phys. Plasmas \textbf{19}, 112502 (2012)]).
By penalizing the island residues in the objective function, the two magnetic field representations are brought into agreement during the optimization.
An example is presented in which, particularly on the surface where quasisymmetry was targeted, quasisymmetry is achieved more accurately than in previously published examples.
\end{abstract}

\maketitle

\section{Introduction}

The geometry of a stellarator can be optimized to improve confinement and stability. Many of the figures of merit that are commonly included in stellarator optimization, such as neoclassical transport and magnetohydrodynamic (MHD) stability, are most conveniently calculated under the assumption that nested toroidal magnetic surfaces exist. Moreover, most of the physics codes in the stellarator community 
use the data format of the Variational Moments Equilibrium Code (VMEC) \cite{VMEC1983}, in which the existence of nested magnetic surfaces is assumed. However, magnetic surfaces are not guaranteed to exist in stellarators, as magnetic islands and chaotic field regions may be present. The extent of islands and chaos is another function of the magnetic geometry that can be optimized\cite{HansonCary,CaryHanson}. Islands and chaos can be diagnosed using other magnetic field representations or MHD equilibrium codes, but then physics objectives that presume the existence of surfaces are not straightforward to calculate.
In stellarator optimization it would therefore be ideal to have the best of both worlds, taking advantage of codes that assume the existence of surfaces and use the VMEC data format, while at the same time controlling islands. In this paper we demonstrate one approach to achieving these goals.

The approach here involves simultaneously using two magnetic field representations, one that assumes the existence of surfaces and one that does not. In particular, we use VMEC and the Stepped Pressure Equilibrium Code (SPEC)\cite{SPEC,SPEC2}, which both compute three-dimensional MHD equilibria. 
VMEC has been widely used in the stellarator community since the 1980s, so a large number of transport and stability codes are available that analyze the numerical solutions it produces. VMEC operates by minimizing the MHD energy subject to two constrained radial profiles and the constraint of nested magnetic surfaces. Hence magnetic islands and chaos cannot be represented. SPEC is a newer code in which the toroidal domain is divided into a number of nested annular regions, with the pressure constant in each region. The magnetic field is constrained to be tangent to the toroidal boundary surfaces of each region, and pressure balance is enforced across these interfaces. Within each of the regions, there is no constraint that magnetic surfaces exist, and so islands and chaos can be represented.
While VMEC has been used as part of stellarator optimization for decades, the present work is the first time SPEC has been used in optimization.

In the new optimization approach presented here, both VMEC and SPEC are run at each  evaluation of the objective function. If any islands are present in the SPEC solution, SPEC and VMEC necessarily will not agree exactly on the magnetic field. A measure of magnetic island width is computed from the SPEC solution and included as a penalty in the objective function. Motivated by Refs.~\onlinecite{HansonCary,CaryHanson}, the measure we use is the residue \cite{Greene}, a real number that can be computed for any periodic field line, which is zero when the island width vanishes. Due to this island width penalty, the islands are eliminated during the optimization so that the VMEC and SPEC representations agree by the end. At the same time, the objective function also includes quantities derived from the VMEC solution. In particular, here we minimize the deviation from quasisymmetry, a continuous symmetry in the field strength that ensures guiding-center confinement\cite{Nuhrenberg}, based on a conversion of the VMEC solution to Boozer coordinates\cite{BoozerCoordinates}. At the start of the optimization, this VMEC-derived part of the objective function may be somewhat inaccurate because the islands were ignored in its computation. But by the end of the optimization, since the islands have been eliminated by the residue penalty, the quasisymmetry term is accurate. In this way, we can take advantage of calculations that are based on the existence of magnetic surfaces, and of the many available codes that postprocess VMEC equilibrium files, while fully accounting for the possibility of magnetic islands.

Compared to earlier island healing calculations done for the National Compact Stellarator Experiment (NCSX) project \cite{Hudson2001, Hudson2002PPCF, Hudson2002PRL, Hudson2003}, there are several difference in the approach here, such as use of different equilibrium codes and physics objectives, and different measures of island width.
Whereas we optimize island widths and other quantities concurrently, the NCSX approach did not use optimization; instead a nonlinear system of equations was solved in which quantities other than island width (such as MHD instability growth rates) were held fixed\cite{Hudson2003}.


\section{Methods}

Here we will consider the first stage of the two-stage optimization procedure used for the design of experiments such as W7-X\cite{Klinger2017} and HSX\cite{Anderson}.
In this first stage, the parameter space for optimization is the shape of a toroidal boundary magnetic surface. Specifically,
the parameter space
consists of the Fourier modes $\{R_{m,n},\,Z_{m,n}\}$ of the boundary
toroidal surface 
\begin{eqnarray}
R(\theta,\phi) &=& \sum_{m,n}R_{m,n}\cos(m\theta-\nfp n\phi), \\ Z(\theta,\phi) &=& \sum_{m,n}Z_{m,n}\sin(m\theta-\nfp n\phi)  \nonumber  
\end{eqnarray}
where $\phi$ is the standard toroidal angle,  $\theta$ is any poloidal angle,  $\nfp$ is the number of field periods, and we have assumed stellarator symmetry. 
We exclude the major radius $R_{0,0}$ from the parameter space, in order to fix the spatial scale.
Here we will not consider the second stage of the two-stage approach, in which coils are optimized to produce the boundary surface resulting from the first stage. In the future, the method of this paper could also be used in a single-stage optimization, in which the parameter space consists of coil shapes, and free-boundary equilibria are used.

For simplicity, we will consider a configuration with no plasma current or pressure. This choice minimizes the computational cost because a single radial domain can be used in SPEC. In the future, the procedure here could be applied to configurations with nonzero plasma current and pressure, using multiple radial domains in SPEC.

The numerical example is carried out using SIMSOPT, a new software framework for stellarator optimization\cite{SIMSOPT,SIMSOPT_Repo}.
The optimization is driven in python, using the default algorithm (trust region reflective)
for nonlinear least-squares optimization from the scipy package\cite{scipy}.
Gradients for the minimization are calculated with forward finite differences,
using MPI for concurrent function evaluations.

The initial state is an axisymmetric circular-cross-section torus, and the number of field periods $\nfp$ is set at two. We
first carry out a preliminary optimization without SPEC or residues. The reason for this preliminary optimization is that the $\iota$ profile evolves significantly at the beginning, causing resonances to enter and leave the domain. In this case it is awkward to include residues in the objective function, because the objective would not be a continuous function of the parameters. The objective function for the preliminary optimization is
\begin{equation}
\label{eq:objective0}
    f =(A-6)^2 + (\iota_0-0.39)^2 + (\iota_a-0.42)^2
     + 2\sum_{m,n} (B_{m,n}/B_{0,0})^2
\end{equation}
where $A$ is the effective aspect ratio as defined in VMEC, $\iota_0$ and $\iota_a$ are the
rotational transform at the magnetic axis and edge, $B_{m,n}$ is the
amplitude of the $\cos(m\vartheta-n\varphi)$ Fourier mode of the field
strength in Boozer coordinates $(\vartheta,\varphi)$ for the flux surface with normalized
toroidal flux $s=0.5$, and only $n \ne 0$ modes are included in the sum. All terms
in the objective are computed from the VMEC solution, with the
$B_{m,n}$ values computed by postprocessing of the VMEC solution with the
BOOZ\_XFORM code\cite{boozxform}.
The aspect ratio is included in the objective because if not, the quasisymmetry term can be reduced to zero by increasing the aspect ratio to infinity.
The rotational transform terms are included in the objective because if there are no constraints on $\iota$, true axisymmetry is an optimum.
The factor of 2 in the quasisymmetry term of (\ref{eq:objective0}) is chosen based on experience to give the best optimum.
Using a script similar to the one described shortly for the combined VMEC-SPEC optimization, the optimization is performed in a series of three steps, as the size of the parameter space and resolution parameters are increased.

For the combined VMEC-SPEC optimization, the objective function is
\begin{eqnarray}
\label{eq:objective}
    f &=&(A-6)^2 + (\iota_0-0.39)^2 + (\iota_a-0.42)^2 \\
    && + 2\sum_{m,n} (B_{m,n}/B_{0,0})^2 + 2R_X^2 + 2R_0^2 \nonumber
\end{eqnarray}
where $R_X$ and $R_O$ are the residues\cite{Greene} for the X- and O-points
of the primary island chain. The residues are
computed from the SPEC solution using Newton's method to find periodic field lines with the desired helicity, then integrating along these field lines to compute the tangent map.
The weight factor of 2 in the residue terms of (\ref{eq:objective}) is chosen by experimentation to yield a good optimum.

\begin{figure}
\includegraphics[width=\columnwidth]{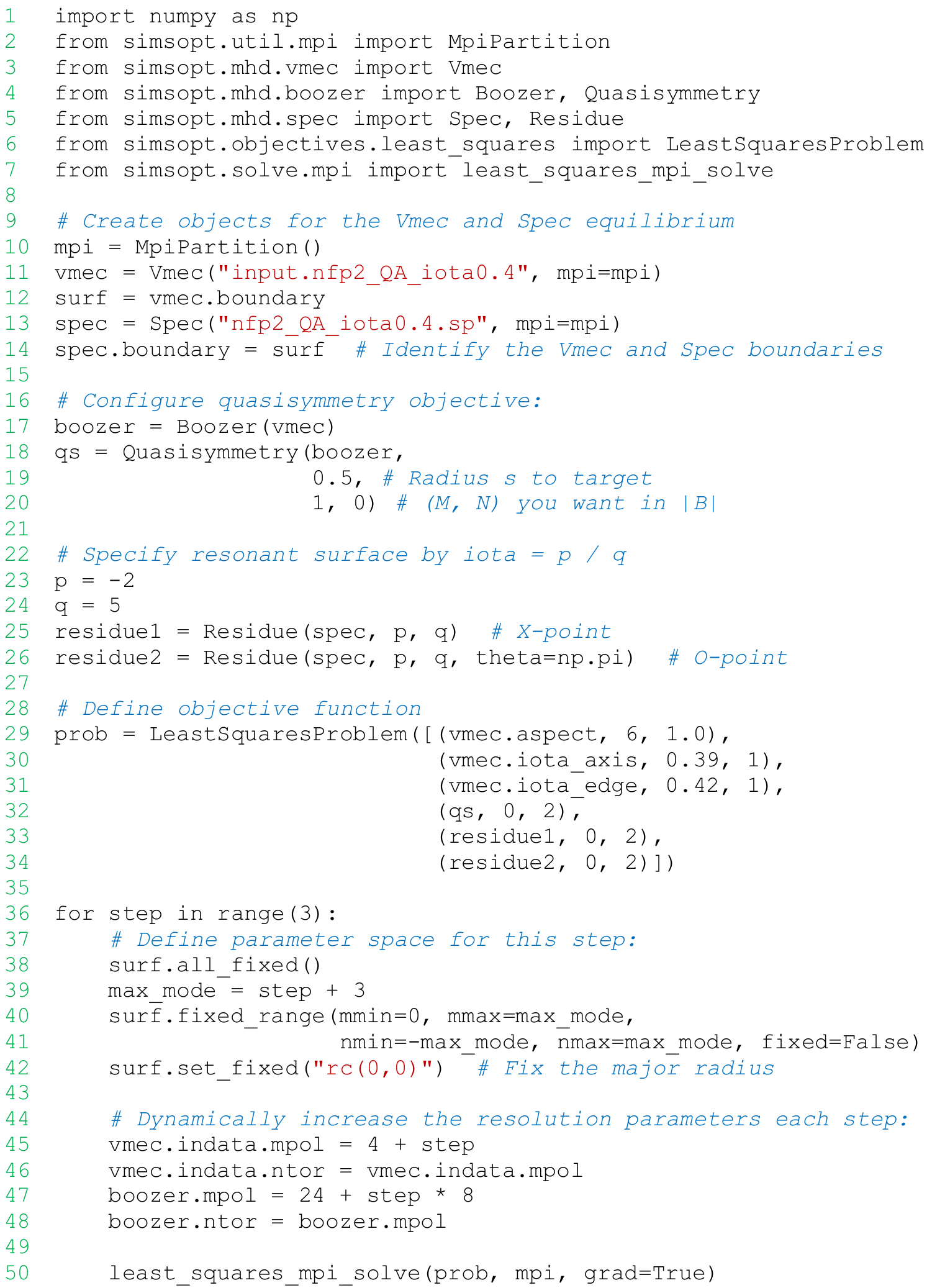}
\caption{\label{fig:code} SIMSOPT driver script for the combined VMEC-SPEC optimization.}
\end{figure}

The SIMSOPT python driver script to define and solve the minimization problem is shown in Fig.~\ref{fig:code}, as several features are noteworthy. On line 14, the \texttt{Spec} object is configured to use the same boundary
\texttt{Surface} object as the \texttt{Vmec} instance. Therefore when the shape of this
single surface is modified during the optimization, the outputs of
both VMEC and SPEC change accordingly.  
The objective function (\ref{eq:objective}) is specified in lines 16-34.
Also, since the optimization
problem is defined with a script, any other desired scripting elements
can be included. Here this capability is used to define a series of
three optimization stages, in which the size of the parameter space
(the maximum $m$ and $n$ values of the $\{R_{m,n},\,Z_{m,n}\}$ to
vary) is increased at each step, along with the numerical resolution
parameters of the codes. The former is valuable to avoid getting stuck
in a poor local minimum, and the latter improves computational efficiency.
In lines 36-41 it can be seen that for the first step, the boundary amplitudes 
$\{R_{m,n},\,Z_{m,n}\}$ are varied for $m = 0 \ldots 3$ and $n = -3 \ldots 3$.
In the second step, the maximum $m$ and $|n|$ are increased to 4, and in the third step, 
the maximum $m$ and $|n|$ are increased to 5. (For the preliminary optimization, the corresponding maximum mode numbers were 1, 2, and 3.) 

\section{Results}

\begin{figure}
\includegraphics[width=\columnwidth]{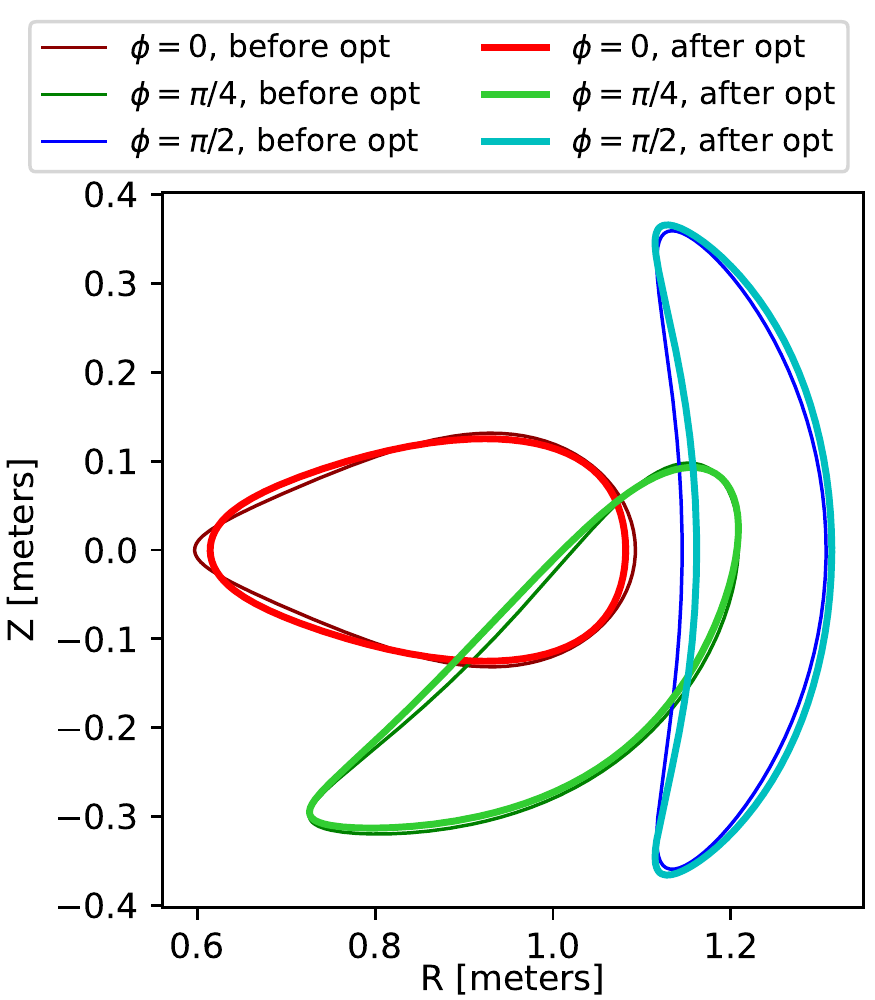}
\caption{\label{fig:xsections} Cross-sections of the plasma before and after the combined VMEC-SPEC optimization.}
\end{figure}

\begin{figure}
\includegraphics[width=\columnwidth]{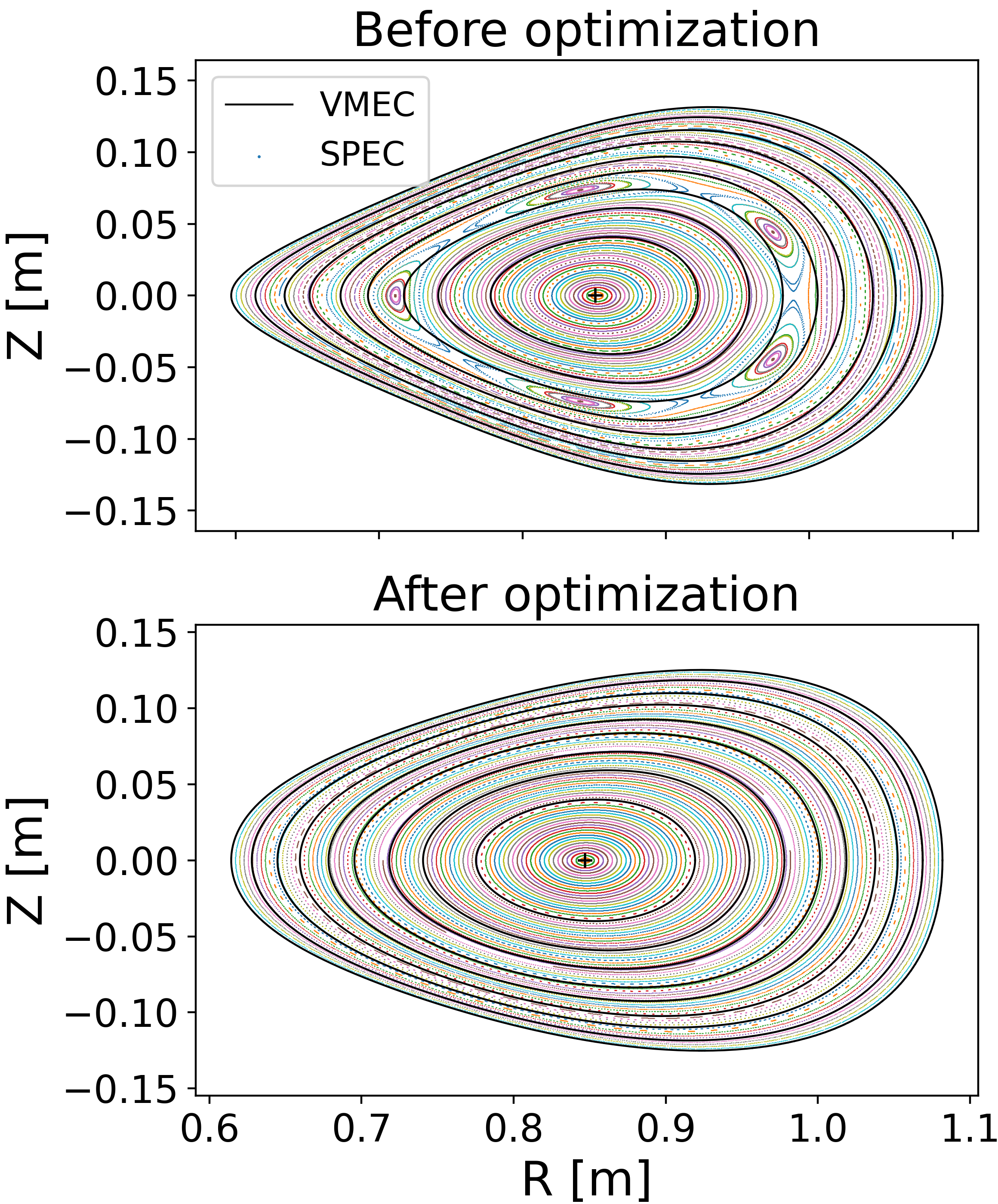}
\caption{\label{fig:poincare} Poincare plots computed from the SPEC solution (colored points), and VMEC magnetic surfaces (black lines). The two codes agree on the surfaces after the combined VMEC-SPEC optimization.}
\end{figure}

\begin{figure}
\includegraphics[width=\columnwidth]{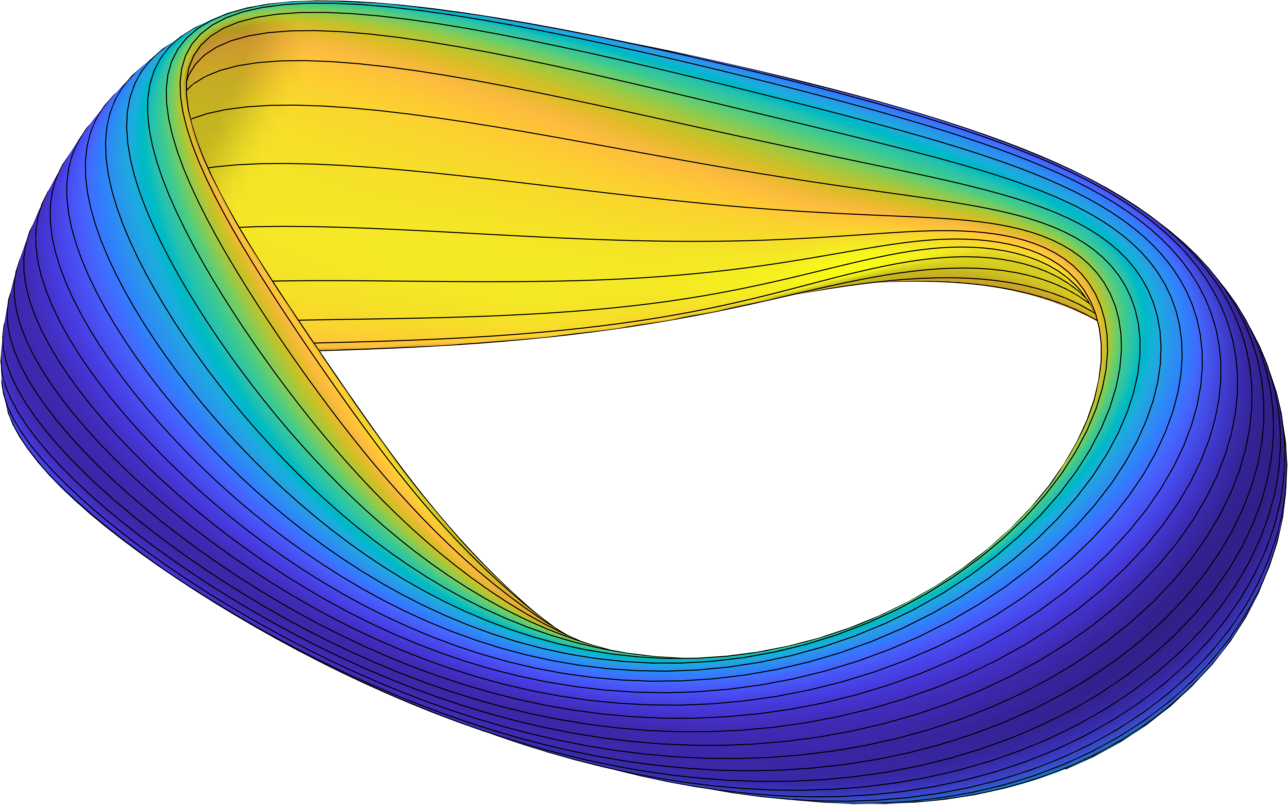}
\caption{\label{fig:3D} The optimized configuration. Color indicates the magnetic field strength, and field lines are shown in black.}
\end{figure}

The configurations before and after the final optimization are shown in
Figs.~\ref{fig:xsections}, \ref{fig:poincare}, and \ref{fig:3D}. 
In the figures, the configuration resulting from the preliminary optimization used as input for the combined VMEC-SPEC optimization is labelled "Before optimization".
Fig.~\ref{fig:poincare} shows this initial configuration has a significant
island chain at the $\iota=2/5$ resonance. Therefore, VMEC and SPEC do
not agree on the internal flux surface shapes near the islands for
this configuration.  
It can also be seen in  Fig.~\ref{fig:poincare} that the optimization
has successfully eliminated the islands. Indeed the magnitudes of the residues have been
reduced from $2\times 10^{-3}$ to $2\times 10^{-6}$. 
The two codes agree very well on the internal surface shapes by the end of the optimization.
Therefore calculations for the final configuration based on the VMEC solution, such as the Boozer coordinate transformation, can be
trusted.

\begin{figure}
\includegraphics[width=\columnwidth]{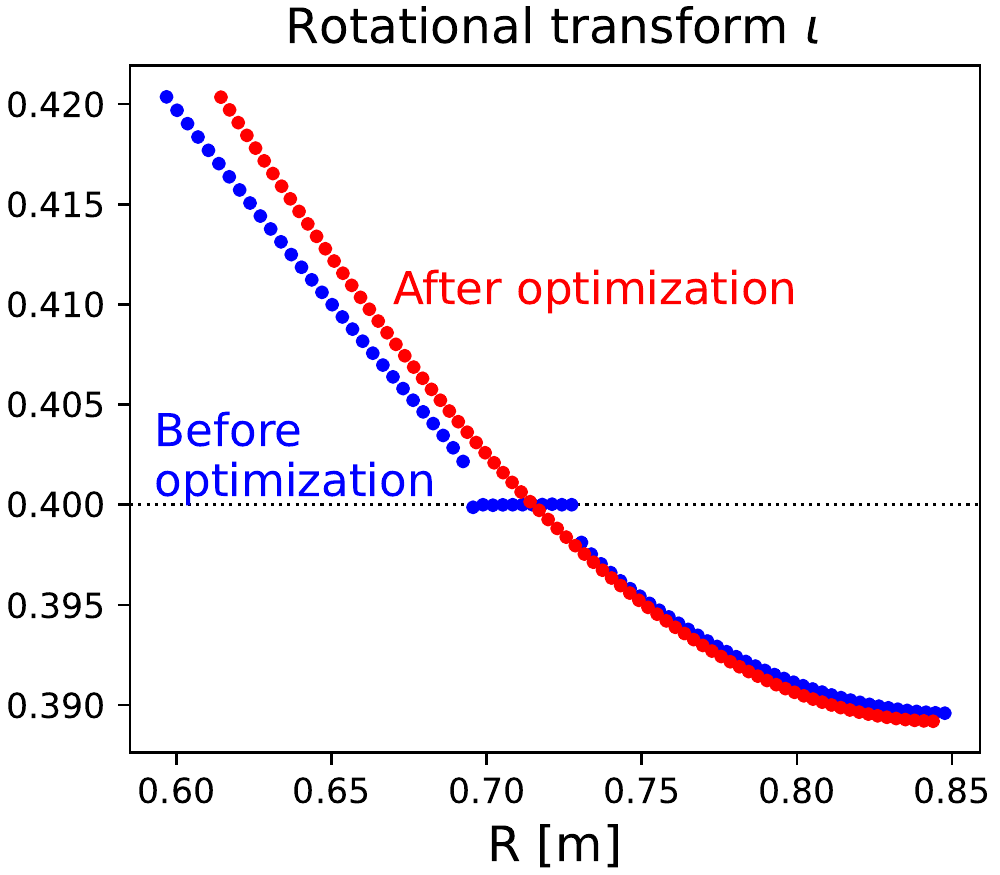}
\caption{\label{fig:iota} Rotational transform profile computed by following field lines in the SPEC solution, before and after the optimization. The islands at $\iota=2/5$ are eliminated even though the rational surface is still in the domain. }
\end{figure}

Fig.~\ref{fig:iota} displays the rotational transform profiles at the beginning and end of the optimization. 
For this figure, $\iota$ was computed by following field lines in the SPEC solution, starting from an array of points on the inboard midplane between the magnetic axis and computational boundary. A flat region in this $\iota$ profile for the initial configuration reflects the presence of a magnetic island.
The $\iota$ profile ranges from 0.39 to 0.42 for both
the initial and optimized configurations, so the islands were not
eliminated by shifting the resonance out of the domain, but rather by
tuning of the resonant field.  

\begin{figure}
\includegraphics[width=\columnwidth]{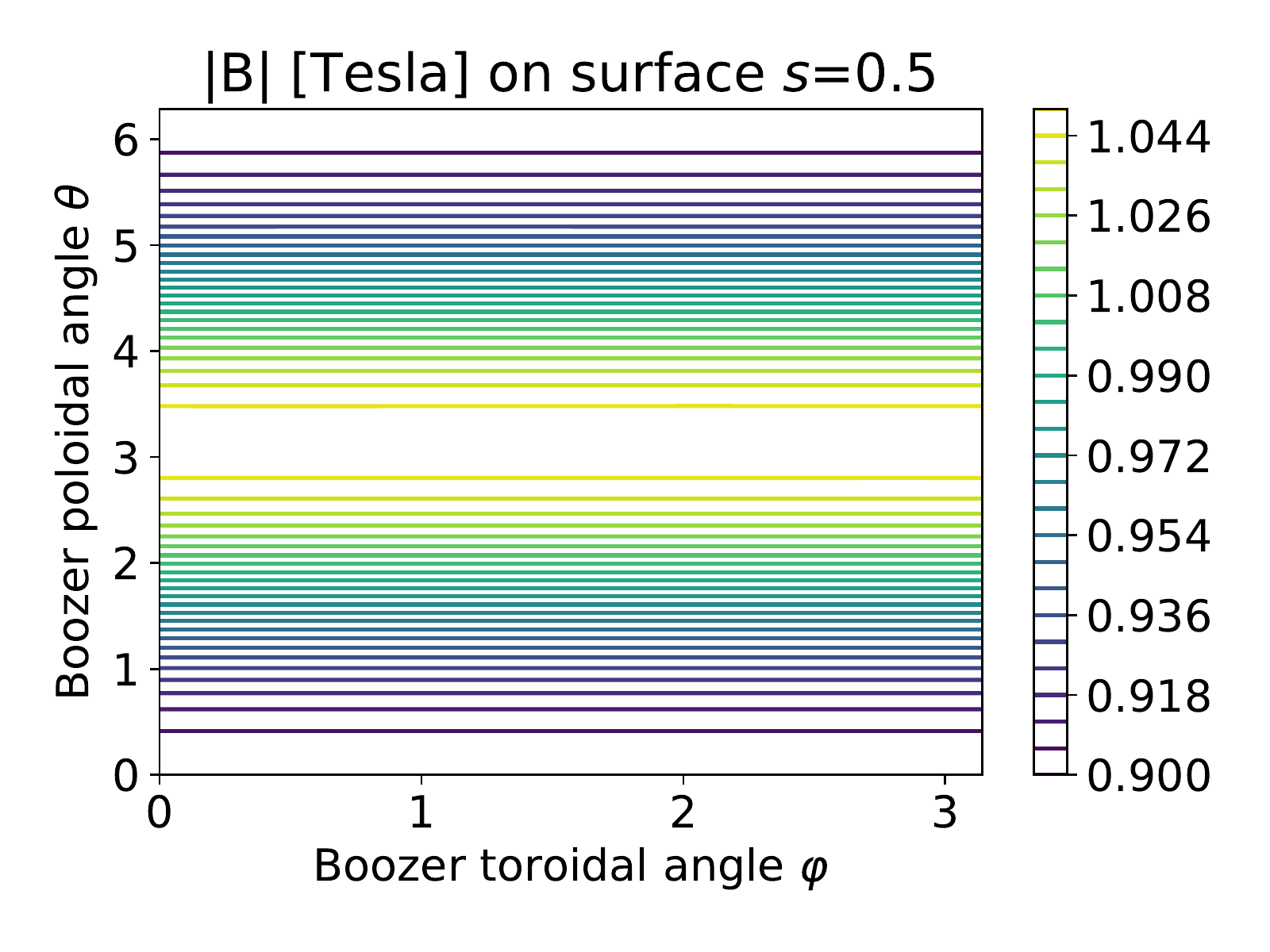}
\caption{\label{fig:boozPlot} Magnetic field strength for the optimized stellarator shape,
 computed from VMEC and BOOZ\_XFORM, showing good
 quasisymmetry.}
\end{figure}

The final configuration also has
extremely good quasiaxisymmetry, especially on the $s=0.5$ surface where symmetry was optimized. This can be seen in the straight horizontal
contours of $|B|$ in Fig.~\ref{fig:boozPlot}.
Any deviation from symmetry is not perceptible in the figure. By contrast, analogous figures of $|B|$ on a surface for previously published quasisymmetric configurations have almost always shown visible ripples in the $|B|$ contours. Examples include Figs.~5-6 of Ref.~\onlinecite{Beidler}, Fig.~2 of Ref.~\onlinecite{CFQS}, and Fig.~4 in Ref.~\onlinecite{Henneberg}. The only previously published configurations we are aware of without clear curvature of the $|B|$ contours are those of Fig.~21 in Ref.~\onlinecite{LandremanSengupta2019}, which are much higher aspect ratio ($\ge 78$).

\begin{figure}
\includegraphics[width=\columnwidth]{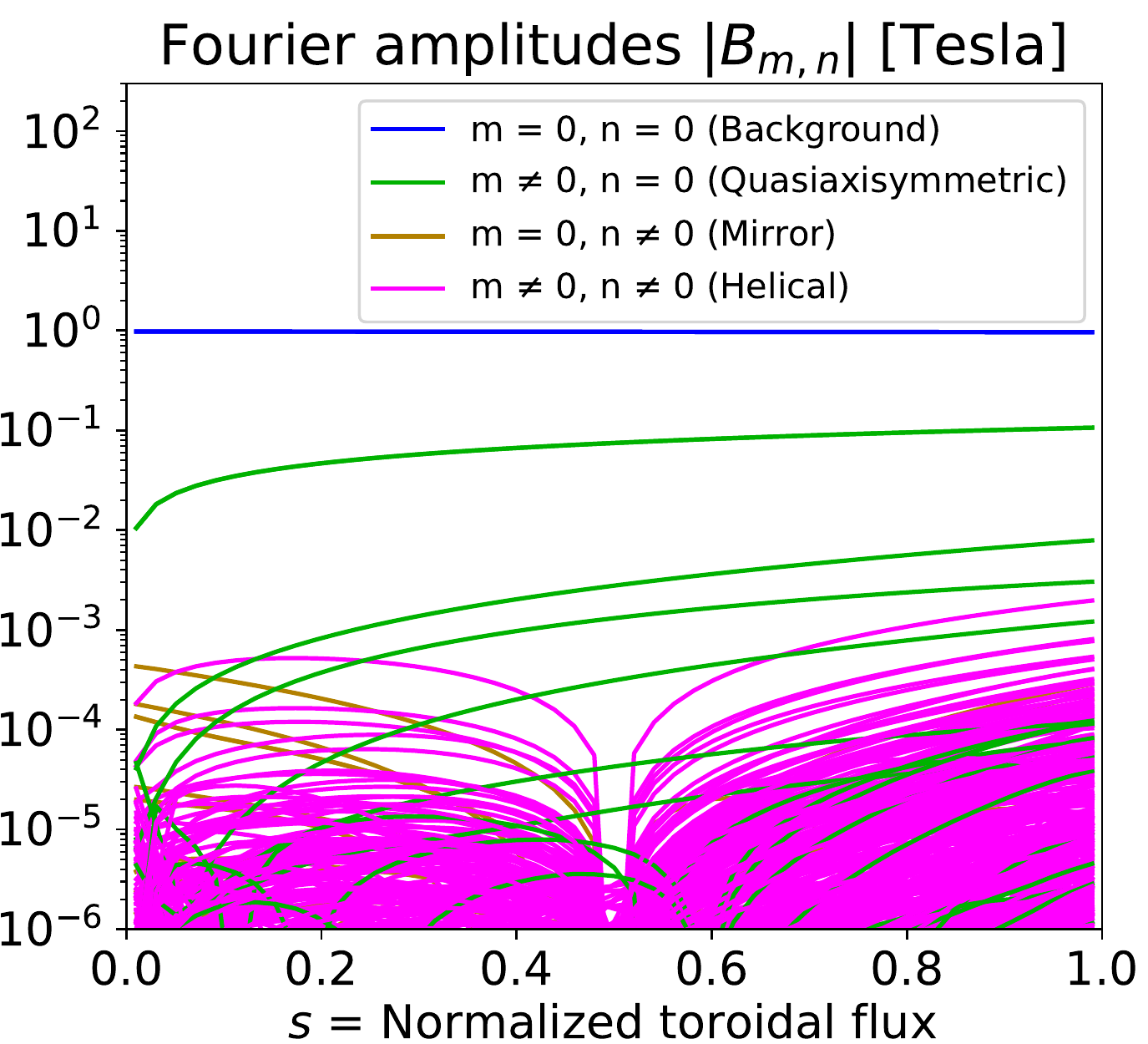}
\caption{\label{fig:boozPlot2} Fourier amplitudes $|B_{m,n}(s)|$ of the magnetic field strength with respect to the Boozer angles for the optimized configuration,
 computed from VMEC and BOOZ\_XFORM. 
The configuration evidently has good quasisymmetry, especially on the the $s=0.5$ surface.}
\end{figure}

The quality of quasisymmetry in the optimized configuration can also be seen in Fig.~\ref{fig:boozPlot2}, which shows the dependence of the $|B_{m,n}|$ amplitudes on minor radius. At $s=0.5$ where symmetry was optimized, there is a striking notch in the $n\ne 0$ modes where their amplitude becomes extremely small, $< 10^{-5}$ of the mean field. This finding supports the conjecture that it may be possible to obtain quasisymmetry exactly on an isolated magnetic surface\cite{GB2}. At the plasma edge, where symmetry-breaking $B_{m,n}$ modes are largest,  the largest nonsymmetric mode remains smaller than all of the symmetric modes for $m=0-3$. 


\section{Discussion}

In this work we have demonstrated
a method for optimizing a stellarator's geometry to eliminate magnetic islands
while simultaneously optimizing other objectives that assume the existence of magnetic surfaces.
This makes it possible for optimizations to include physics codes that use equilibria from the VMEC code, or other equilibrium codes that assume the existence of nested magnetic surfaces, while also ensuring good surface quality.
This method can be applied in the future to configurations with nonuniform pressure by using multiple radial domains in SPEC.
While quasi-axisymmetry was the main objective in the example here, the method is equally applicable to other objectives.

Another extension of this work could be to use a measure of flux surface quality or island width other than residues. In principle any such method could be used in the procedure of this paper. 
Several such alternative measures include 
Mather's\cite{Mather,Meiss} $\Delta W$, the estimate by Cary and Hanson\cite{CaryHansonIslandWidth},
and converse KAM\cite{Meiss,converseKAM}.

\begin{acknowledgments}
We gratefully acknowledge discussions with and assistance from Aaron Bader, David Bindel, Benjamin Faber, Andrew Giuliani, Stuart Hudson, Rogerio Jorge, Thomas Kruger, Zhisong Qu, Jonathan Schilling, Florian Wechsung, and Mike Zarnstorff.
This work was supported by a grant from the Simons Foundation (560651, ML).
BM and CZ are supported by the U.S. Department of Energy under
Contract No. DE-AC02-09CH11466 through the Princeton Plasma Physics
Laboratory.
\end{acknowledgments}

\section*{Data Availability Statement}

The data that support the findings of this study are openly available in Zenodo at \url{http://doi.org/10.5281/zenodo.5035515}.



\bibliography{combined_vmec_spec_optimization}

\end{document}